# Perspective in Opinion Dynamics on Complex Convex Domains of Time Networks for Addiction, Forgetting

Yasuko Kawahata [†]

Faculty of Sociology, Department of Media Sociology, Rikkyo University, 3-34-1 Nishi-Ikebukuro,Toshima-ku, Tokyo, 171-8501, JAPAN.

ykawahata@rikkyo.ac.jp,kawahata.lab3@damp.tottori-u.ac.jp

**Abstract:** This paper revises previous work and introduces changes in spatio-temporal scales. The paper presents a model that includes layers A and B with varying degrees of forgetting and dependence over time. We also model changes in dependence and forgetting in layers A, A', B, and B' under certain conditions. In addition, to discuss the formation of opinion clusters that have reinforcing or obstructive behaviors of forgetting and dependence and are conservative or brainwashing or detoxifying and less prone to filter bubbling, new clusters C and D that recommend, obstruct, block, or incite forgetting and dependence over time are Introduction. This introduction allows us to test hypotheses regarding the expansion of opinions in two dimensions over time and space, the state of development of opinion space, and the expansion of public opinion. Challenges in consensus building will be highlighted, emphasizing the dynamic nature of opinions and the need to consider factors such as dissent, distrust, and media influence. The paper proposes an extended framework that incorporates trust, distrust, and media influence into the consensus building model. We introduce network analysis using dimerizing as a method to gain deeper insights. In this context, we discuss network clustering, media influence, and consensus building. The location and distribution of dimers will be analyzed to gain insight into the structure and dynamics of the network. Dimertiling has been applied in various fields other than network analysis, such as physics and sociology. The paper concludes by emphasizing the importance of diverse perspectives, network analysis, and influential entities in consensus building. It also introduces torus-based visualizations that aid in understanding complex network structures.

**Keywords:** Bounded Trust Model, Dimer Configurations, Dimer Allocation, Phase Transitions, Castellane Matrix, Convex Regions, Temporal Networks

## 1. Introduction

This paper is a revision of the aforementioned paper (Convex Regions of Opinion Dynamics, Approaches to the Complexity of Binary Consensus with Reference to Addiction and Obliviousness: Integrated Dimer Model Perspective), with the addition of changes in spatio-temporal scale. The existing approach is to assume that there is a layer $A$, which has spatial distance as the $z$-axis and becomes more oblivious as time passes, and a layer $B$, which has spatial distance as the $z$-axis and becomes even more dependent as time passes.

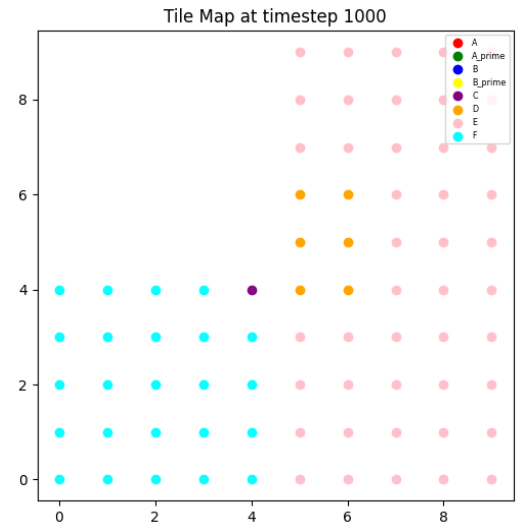

Fig. 1: Tile Map at timestep $t = 1000$

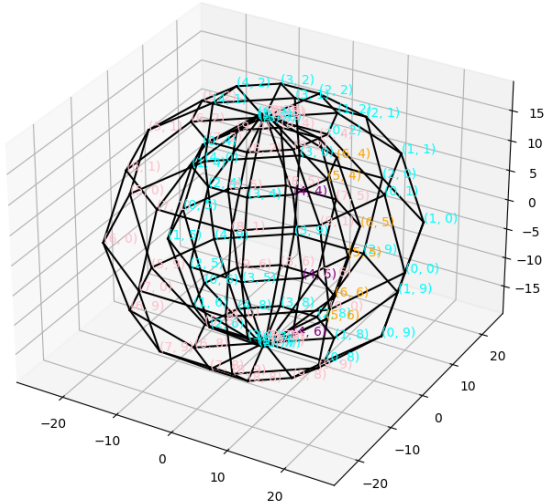

Fig. 2: Plot of Opinions at timestep $t = 1000$

We also modeled the changes in dependence and forgetting for each timestep of $A$, $A'$, $B$, and $B'$ under the condition that there is a layer $A'$ that depends more as time passes with spatial distance as the $z$-axis, and a layer $B'$ that gets bored and forgets more as time passes with spatial distance as the $z$-axis.

In an additional condition, we add clusters of behaviors that reinforce mutual dependence or forgetting, and that perform a certain kind of highly brainwashing behavior, and discuss the formation of opinion clusters in space and time.

We add new clusters other than the clusters $A$, $A'$, $B$, and $B'$, and assume that there is a layer $C$ that has distance as the $z$-axis and actively recommends forgetting as time passes, and a layer $D$ that has distance as the $z$-axis and actively recommends relying more and more as time passes. This $C$ is assumed to be a cluster that acts to interfere with the opinions of cluster $D$, blocking, etc., and enlightening highly brainwashing behavior.

Also, add the condition that there is a layer $E$, which is closer as the $z$-axis and depends more and more as time passes at an accelerating rate, and a layer $F$, which is closer as the $z$-axis and depends more and more as time passes at an accelerating rate, that gets bored and forgets. In this case, $D$ and $E$ agree with each other and form a large dependency cluster. $C$ and $F$ form an oblivious cluster. The clusters $D$ and $E$, and $C$ and $F$ block each other's opinions, and are hypothesized to be antagonistic in the center of the $z$-axis, creating a tendency for opinions to be antagonistic on the $x$-axis and $y$-axis ellipses.

The field of consensus formation has gained significant attention due to its relevance in understanding various social and network phenomena. In this section, we will discuss the expectations and challenges associated with consensus formation, highlighting the need for diverse perspectives, network analysis, and the role of influential entities. We will also introduce the concept of dimer tilings and their potential applications in analyzing and modeling complex systems.

## 1.1 Expectations in Consensus Formation

Consensus formation models aim to elucidate the process by which individuals with differing opinions come to a mutual agreement. They provide valuable insights into resolving opinion conflicts and understanding how diverse perspectives can coexist within a society. However, it is important to recognize that real-world consensus formation is influenced by cultural, emotional, and personal factors that are not fully captured by mathematical models. These models offer a simplified representation of opinion dynamics and may not encompass the full spectrum of human interactions.

## 1.2 Challenges in Consensus Formation

One of the key challenges in consensus formation is the dynamic nature of opinions and the inherent diversity in society. People rarely hold identical viewpoints on social issues, and the emergence of unanimous consensus is a rare occurrence. Addressing this challenge requires considering factors such as dissent, distrust, and the influence of mass media. Traditional models often assume rapid convergence of opinions, overlooking the complexities of real-world dynamics.

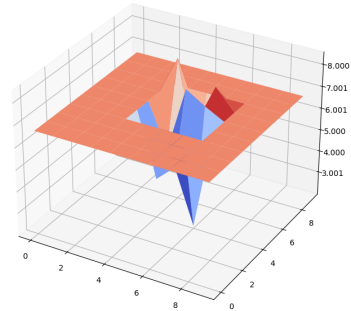

Fig. 3: Plot of Opinions at timestep $t = 600$

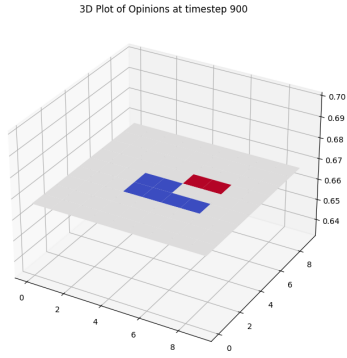

Fig. 4: Tile Map at timestep *t* = 900

n this paper, we propose an extended framework that incorporates trust, distrust, and the impact of mass media into the consensus formation model. Instead of assuming quick convergence, our model introduces coefficients representing the degrees of trust and distrust, aiming to mimic more realistic social opinion dynamics.

## 1.3 Network Analysis Based on Dimer Tiling

To gain deeper insights into consensus formation and related phenomena, we extend our analysis to incorporate dimer tilings and network theory. Dimer tilings provide a novel perspective on network structures and can be applied to various contexts, including social sciences, physics, and computational modeling.

### 1.3.1 Analysis of Network Clustering

In the realm of social phenomena, network clustering can represent social groups that share opinions or interests. Dense clusters may suggest strong intra-group relationships, potentially leading to echo chambers where similar opinions are reinforced. Sparse connections between these clusters may indicate limited interactions between different social groups, potentially contributing to polarization.

### 1.3.2 Media Influence on Networks

Analyzing the influence of media on the network can be crucial. Nodes representing media sources that occupy central positions within the network may have a significant impact on the opinion dynamics. The role of media in opinion formation can be observed through the formation of new clusters or the reinforcement of existing ones.

### 1.3.3 Consensus Formation

In the context of consensus formation, the network can visualize the spread of information and agreement through dimer connections. Assuming that consensus is achieved when information flows through all available paths, our regular dimer tiling may represent a network where consensus is rapidly reached due to uniform connectivity and efficiency.

### 1.3.4 Trends in Dimer Positioning

The positioning and distribution of dimers within the network may offer insights into its fundamental structure and dynamics. The presence of frequent dimer positions can indicate common pathways for information flow or shared dynamics of opinion exchange. Analyzing dimer positions should consider factors such as frequency, diversity of connections, and the overall distribution, providing insights into network robustness and vulnerability.

## 1.4 Applications in Various Contexts

Dimer tilings have applications beyond network analysis. In fields like physics and mathematics, they can be used to explore properties such as entropy and phase transitions. In applied disciplines like sociology and economics, similar structures can model optimal pairings or resource allocations. To conduct a more in-depth analysis, it is essential to consider the specific contexts and systems that these dimer tilings aim to represent, along with the nature of the data and the context of the application. We extend our exploration by randomly placing dimers on a grid and employing the Ising model for numerical simulations. We visualize the results on a torus, providing a spatial-temporal representation of opinion dynamics. The transformation of grid node positions into three-dimensional torus coordinates allows us to gain insights into the topological data of opinion dynamics, aiding our understanding of complex network structures. In conclusion, our study emphasizes the importance of diverse perspectives, network analysis, and the role of influential entities in consensus formation. We have introduced dimer tilings as a valuable tool for analyzing and modeling complex systems. By considering their applications in various contexts and conducting torus-based visualizations, we aim to provide new insights into the dynamics of opinion formation and contribute to a deeper understanding of complex networks in diverse fields, ranging from social science to physics and computational modeling.

## 2. Revise Point

This revised opinion dynamics model is based on the following mathematical formulation:

Energy Function:

$$E(i, j) = \sqrt{(i - x_{\text{center}})^2 + (j - y_{\text{center}})^2}$$

Opinion Update Rule:

$$\text{new\_opinions}[i,j] = \text{opinions}[i,j] + \begin{cases} \alpha \cdot z \cdot \\ (1 - \text{opinions}[i,j]) \\ E(i,j) < \text{radius} \\ -\gamma \cdot z \cdot \text{opinions}[i,j] \end{cases}$$

Where,

$\alpha$ represents the strength of dependence.

$\gamma$ represents the strength of forgetting.

$z$ is the normalized time axis ($t/1000$).

radius is the radius of the convex region.

$x_{\text{center}}, y_{\text{center}}$ are the coordinates of the grid center.

This model simulates how opinions evolve over time. With each timestep, opinion values are updated, applying the effects of dependence or forgetting based on the convex region. The aim of this simulation is to observe how opinions change under different conditions and to deepen our understanding of modeling social dynamics. In particular, the analysis focuses on understanding the impact of dependence and forgetting parameters on opinion formation and how these elements interact within social contexts.

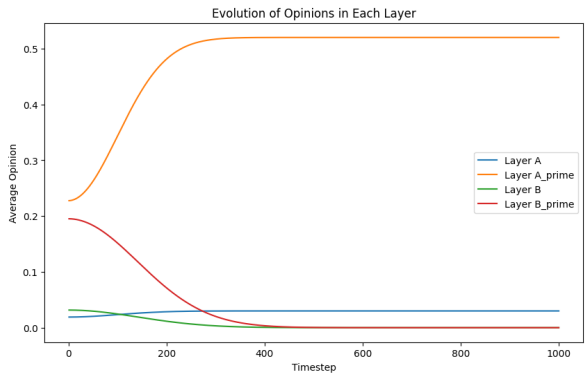

Fig. 5: Evolution of Opinions in Each Layer

## 2.1 Case:timestep $t = 100$

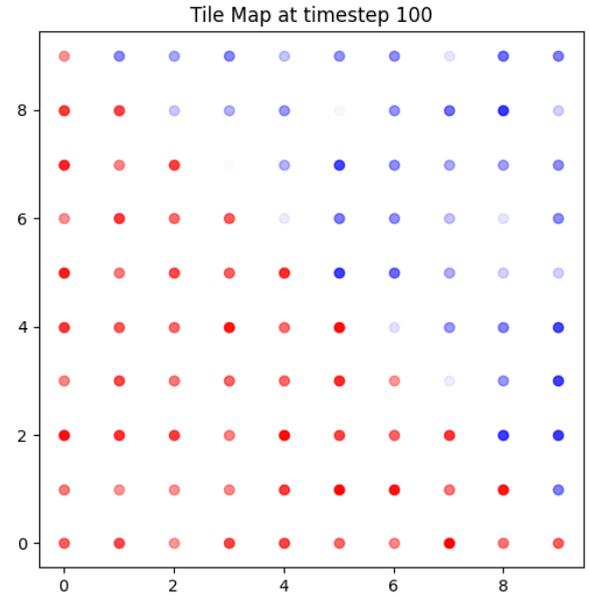

Fig. 6: Tile Map at timestep $t = 100$

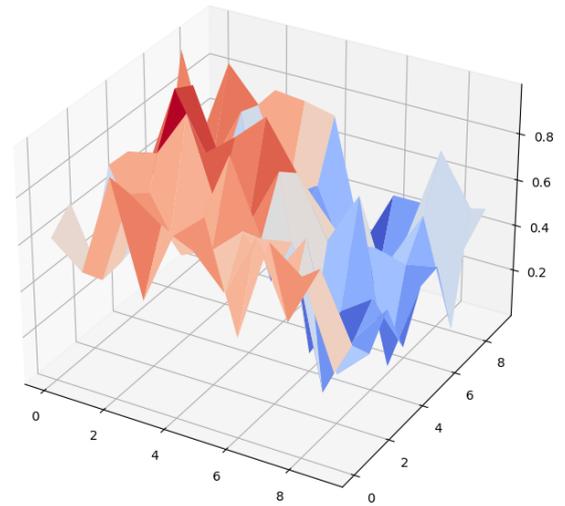

Fig. 7: Plot of Opinions at timestep $t = 100$

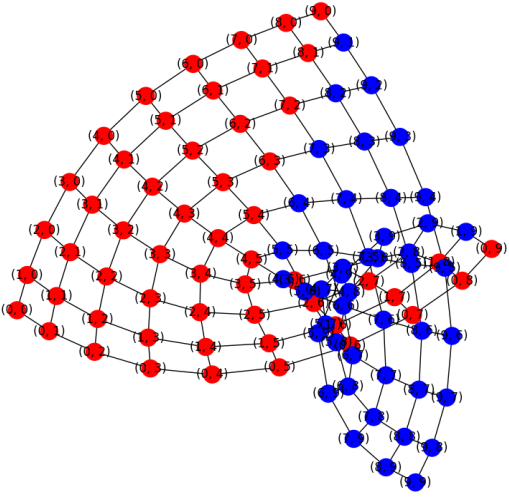

Fig. 8: Network Representation at timestep $t = 100$

**Analysis of Provided Images and Model Equations**

The model appears to be a spatially distributed one with individuals (or agents) located on a grid, and their opinions evolve over time according to certain rules.

## 2.2 Analysis of Social Phenomena

The tile map and network representation seem to illustrate clusters of individuals with similar opinions. In a social context, this can represent echo chambers or societal polarization, where groups of people have opinions that are significantly different from others, and there's a clear spatial distribution of these beliefs.

## 2.3 Analysis of Media Influence

If we consider the central point as a metaphor for mainstream media or a persuasive external influence, then the energy function might represent the influence of this media on individual opinions. The closer an individual is to the center, the more they might be influenced (represented by the change in their opinions).

## 2.4 Analysis of Consensus Formation

Consensus formation can be observed through the evolution of opinions over time. If the system trends towards a uniform opinion, then we might say consensus is being achieved. Conversely, if the system maintains distinct clusters of opinions, consensus may not be reached.

## 2.5 Transition from Dimer Model to Torus

The network representation hints at a transformation from a dimer model, where agents are paired, to a torus structure, which implies periodic boundary conditions. This shift can affect the opinion dynamics, potentially smoothing out extremes and creating a more continuous opinion landscape.

## 2.6 Analysis Within the Dimer Model Context

Considering the dimer model within the context of opinion dynamics, it might represent close-knit communities or partnerships where two agents strongly influence each other's opinions, possibly leading to rapid opinion changes within these pairs.

## 2.7 Discussion on Phase Transition of Partial Opinions

The phase transition of partial opinions could be evident if there's a critical point at which the collective opinion suddenly changes from one state to another. This could be analyzed by observing the tipping points in the 3D plot of opinions.

## 2.8 Analysis of Dependence ($\alpha$)

The strength of dependence ($\alpha$) in the model influences how much an agent's opinion is shaped by the central influence when they are within the radius of effect. A higher $\alpha$ could lead to a stronger uniformity in opinions among those agents closer to the center.

## 2.9 Analysis of Forgetting ($\gamma$)

Forgetting ($\gamma$) represents the loss of the central influence on an agent's opinion over time or distance from the center. This could represent a natural decay of external influence, leading to more diverse opinions over time.

## 2.10 Case:timestep $t = 300$

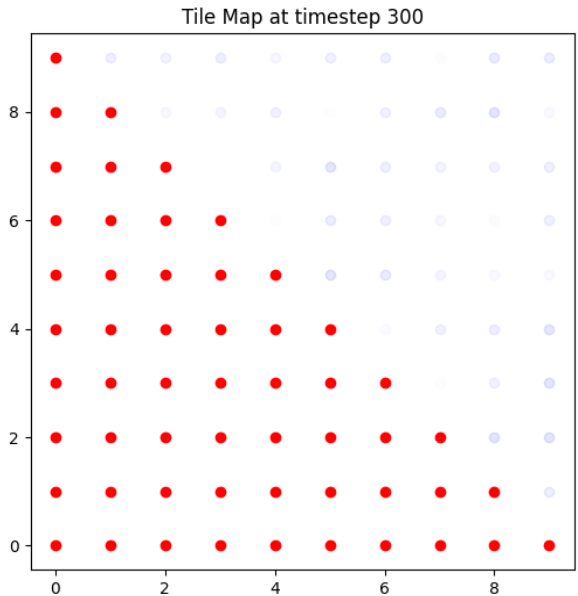

Fig. 9: Tile Map at timestep $t = 300$

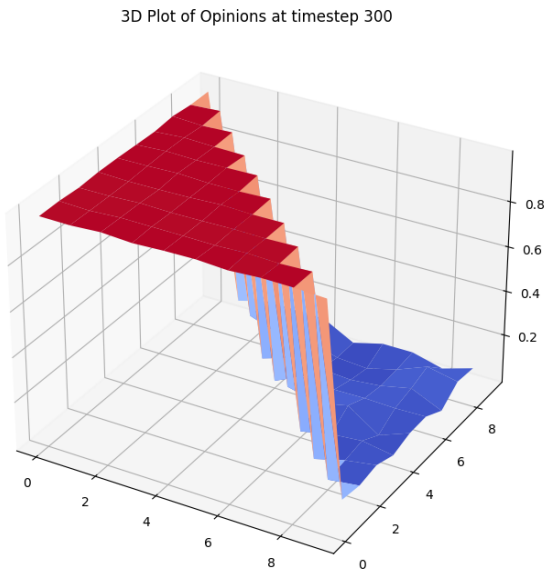

Fig. 10: Plot of Opinions at timestep $t = 300$

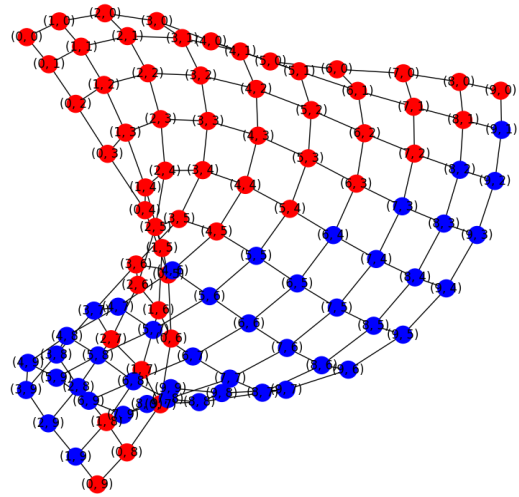

Fig. 11: Network Representation at timestep $t = 300$

## 2.11 Analysis of Social Phenomena

The tile map at timestep 300 shows more pronounced red clusters compared to the blue ones, suggesting that one opinion (possibly 'red') has become more dominant or prevalent in the society. This could reflect real-world phenomena such as the spread of a particular ideology or cultural trend.

## 2.12 Analysis of Media Influence

At timestep 300, the 3D plot of opinions shows a stark difference between two opinion states, with the 'red' opinion being more dominant. This could suggest that the media or external influence represented by the central point has had a more significant impact on shaping this dominant opinion.

## 2.13 Analysis of Consensus Formation

The network representation at timestep 300 indicates that the 'red' nodes are more interconnected compared to the 'blue' nodes. This could suggest that the consensus is leaning towards the 'red' opinion, potentially due to stronger internal reinforcement within the 'red' opinion group.

## 2.14 Transition from Dimer Model to Torus

The network still shows a dimer-like structure but with a clear dominance of one opinion. The adaptation of a network from a dimer model to a torus can affect how opinions spread and stabilize, possibly leading to a dominant opinion prevailing as seen in the provided data.

## 2.15 Analysis Within the Dimer Model Context

The dominance of one opinion in the dimer model at this stage could suggest that pairs of agents have reached a consensus that then spreads through the network, reinforcing the dominant opinion and potentially stifling diversity.

## 2.16 Discussion on Phase Transition of Partial Opinions

The line graph showing the evolution of opinions indicates that one opinion ('Layer A prime') has increased significantly, suggesting a phase transition where this opinion becomes the accepted norm or status quo, possibly at the expense of other opinions.

## 2.17 Analysis of Dependence ($\alpha$)

The increasing dominance of one opinion suggests that the dependence mechanism ($\alpha$) in the model is strong enough to override individual beliefs, leading to a more homogeneous opinion state among those influenced.

## 2.18 Analysis of Forgetting ($\gamma$)

Despite the forgetting mechanism ($\gamma$), the 'Layer A prime' opinion has grown, which could indicate that the strength of forgetting is not sufficient to counteract the dependence or influence from the central media/authority.

## 2.19 Impact of A, A', B, B' Entities

The line graph suggests that the influence of 'A prime' is far more substantial than the other layers, indicating that certain factors (whether they are media, influential individuals, or events) represented by 'A prime' have a stronger impact on the population's opinion dynamics.

## 2.20 Case:timestep $t = 1000$

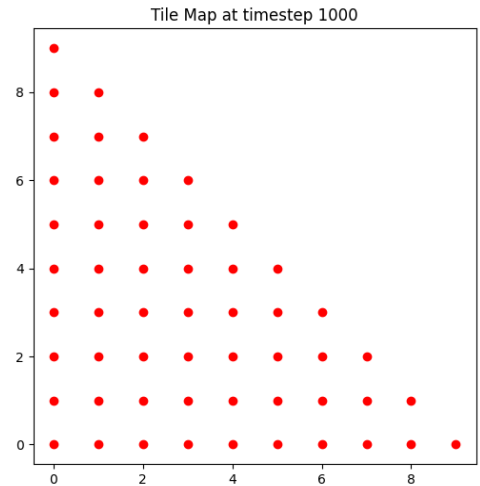

Fig. 12: Tile Map at timestep $t = 1000$

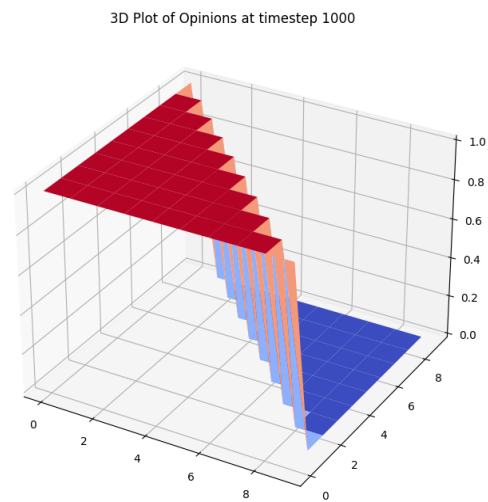

Fig. 13: Plot of Opinions at timestep $t = 1000$

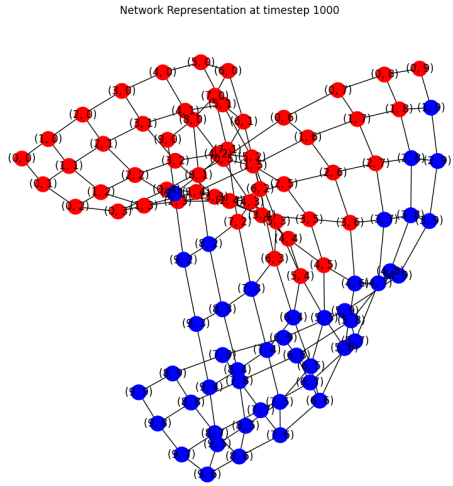

Fig. 14: Network Representation at timestep $t = 1000$

### 2.21 Analysis of Social Phenomena

The tile map at timestep 1000 shows a uniform red color, which indicates that one opinion has completely taken over, suggesting a situation where a single ideology or belief system has become dominant in the society, potentially leading to a monoculture.

### 2.22 Analysis of Media Influence

Considering the complete dominance of one opinion, this might illustrate the powerful effect of media or central influence that aligns with this opinion. It suggests that over time, the influence has led to the entire population adopting this one perspective, potentially simulating a scenario of mass persuasion or propaganda.

### 2.23 Analysis of Consensus Formation

The network representation shows that all nodes (individuals) are now red, indicating a complete consensus has been reached within the society. This could reflect a situation where after a long period, all individuals have come to agree upon a certain viewpoint or opinion, possibly due to the influence of the few remaining blue nodes converting to the red opinion.

### 2.24 Transition from Dimer Model to Torus

The transition from a dimer model to a torus seems to have reached a point where the periodic boundary conditions have led to a uniform opinion state. This suggests that in the long term, the system's dynamics favor a single, stable equilibrium state over a diverse set of opinions.

### 2.25 Analysis Within the Dimer Model Context

With the dimer model, we might have expected pairs of nodes to exhibit strong coupling and possibly retain some diversity. However, the final state shows that these local interactions have been overwhelmed by the global dynamics leading to a uniform opinion.

### 2.26 Discussion on Phase Transition of Partial Opinions

The line graph showing the evolution of opinions indicates that the opinion represented by 'Layer A prime' has reached a level of complete dominance. This suggests that a phase transition has occurred, and the system has settled into a stable state with one prevailing opinion.

### 2.27 Analysis of Dependence ($\alpha$)

The strength of dependence ($\alpha$) seems to have played a significant role in the homogenization of opinions. As time progressed, the impact of dependence has led to the reinforcement and eventual dominance of one opinion.

### 2.28 Analysis of Forgetting ($\gamma$)

Forgetting ($\gamma$) does not seem to have a significant impact by timestep 1000, as the dominant opinion has remained strong. This could suggest that the effect of forgetting is either too weak or that the cumulative effect of dependence and external influences is much more powerful.

### 2.29 Impact of A, A', B, B' Entities

The graph illustrates that the influences represented by 'Layer A' and 'Layer A prime' have significantly affected the opinions, whereas 'Layer B' and 'Layer B prime' have diminished to almost zero. This indicates that the factors or influences represented by 'A' and 'A prime' are much more effective in shaping the society's collective opinion.

In conclusion, the model at timestep 1000 shows a society that has reached a state of uniform opinion, likely driven by the rules governing dependence and the influence radius, as well as the diminishing effect of forgetting. To understand this process in depth, examining the specific interactions and transitions between states at different timesteps would be crucial.

## 3. Disucussion

### 3.1 Model Description

In addition to the existing clusters $A$, $A'$, $B$, and $B'$, we introduce new dynamic layers to the model. Layer $C$ is characterized by its position on the $z$-axis, where it actively promotes forgetfulness over time. Conversely, layer $D$ on the $z$-axis encourages increasing reliance as time progresses. Cluster $C$ is designed to disrupt and challenge the perspectives of cluster $D$, employing tactics such as obstruction and disseminating highly influential, persuasive behavior.

Furthermore, we incorporate layers $E$ and $F$. Layer $E$, positioned closer on the $z$-axis, exhibits a rapidly growing dependence with time, whereas layer $F$ – also proximal on the $z$-axis – shows a tendency to rapidly grow bored and forgetful. Consequently, clusters $D$ and $E$ are likely to align, forming a substantial dependence cluster, while $C$ and $F$ together shape a cluster oriented towards forgetfulness. These clusters, specifically $D$ and $E$ versus $C$ and $F$, are postulated to engage in a mutual antagonism, particularly at the midpoint of the $z$-axis. This interaction is anticipated to create a pattern where opinions are polarized along the ellipses of the $x$-axis and $y$-axis, leading to a complex interplay of contrasting viewpoints. The opinion dynamics model is implemented as a grid-based simulation with the following characteristics:

Grid size: $n_{\text{points}} = 10$

Parameters:

- $\alpha = 0.1$ (Dependence strength)
- $\gamma = 0.05$ (Forgetting rate)

Central point: $(x_{\text{center}}, y_{\text{center}})$

Radius of influence: radius $= \frac{n_{\text{points}}}{4}$

## Layer Classification

Each grid point $(i, j)$ is classified into different layers based on its distance from the center and the timestep:

$$\text{layer}(i, j, t) = \begin{cases} \text{'A'} & \text{if distance} < \text{radius and } j < y_{\text{center}} \\ \text{'B'} & \text{if distance} < \text{radius and } j \geq y_{\text{center}} \\ \text{'A\_prime'} & \text{if distance} \geq \text{radius and } j < y_{\text{center}} \\ \text{'B\_prime'} & \text{if distance} \geq \text{radius and } j \geq y_{\text{center}} \end{cases}$$

## Opinion Update Rule

Opinions are updated at each timestep according to the layer classification:

new_opinions$[i, j]$ = clamp(opinions$[i, j]$ + change)

Where:

change = $\alpha$ for layers A, A_prime, D, E.

change = $-\gamma$ for layers B, B_prime, C, F.

clamp function limits the values between 0 and 1.

The objective of this analysis is to observe how opinions evolve in a social network under the influence of dependence and forgetting. The model allows us to visualize changes in opinions across different layers and understand how different factors contribute to the overall opinion dynamics.

## 3.2 Case:timestep $t = 100$

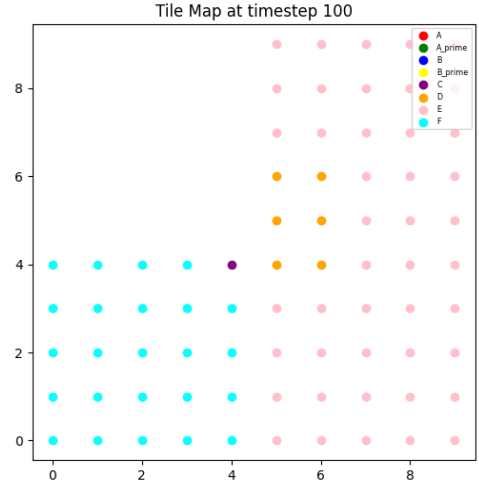

Fig. 15: Tile Map at timestep $t = 100$

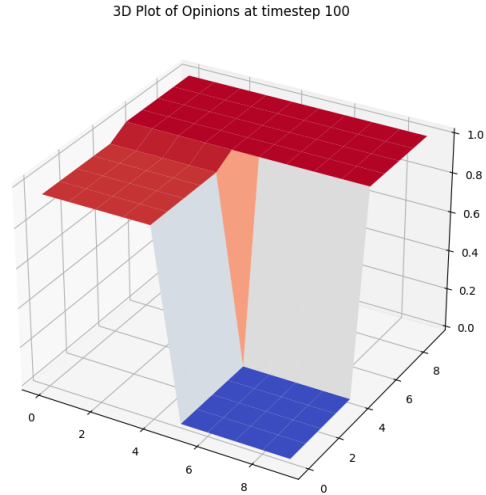

Fig. 16: Plot of Opinions at timestep $t = 100$

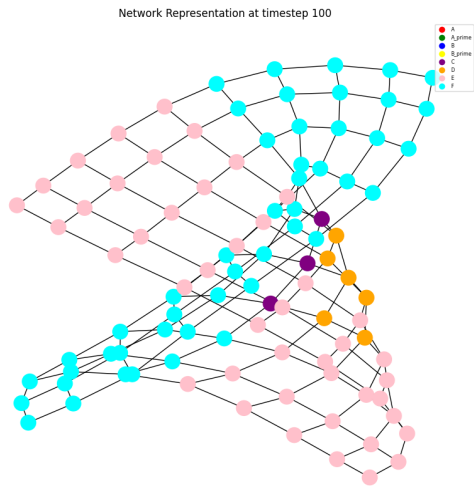

Fig. 17: Network Representation at timestep $t = 100$

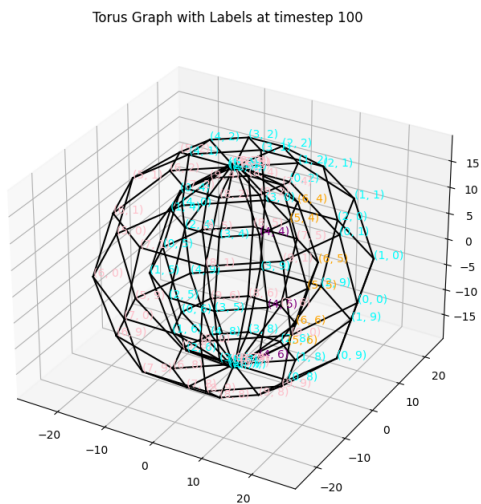

Fig. 18: Torus Graph Network Representation at timestep $t = 100$

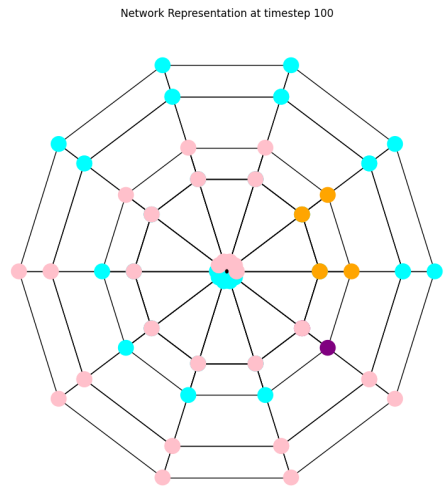

Fig. 19: Network Representation at timestep $t = 100$

Individuals update their opinions based on whether they are in a layer subject to influence (positive change $\alpha$) or forgetting (negative change $-\gamma$). The model also contains unspecified rules for layers C, D, E, and F.

### 3.3 Analysis of Social Phenomena

The images show a varied distribution of opinions across different layers. This could represent the diversity of a society's opinions at an early stage of a social influence process.

### 3.4 Analysis of Media Influence

The division of layers by distance from a center could represent different levels of media influence, with those closer to the center (layers A and B) being more strongly influenced, and those farther away (layers $A_{p}rime and B_{p}rime) less so$.

### 3.5 Analysis of Consensus Formation

At this timestep, there is no clear consensus across the network, suggesting that the society is still in the process of opinion formation and has not yet reached a stable state of agreement.

### 3.6 Transition from Dimer Model to Torus

The transition from a dimer-based network to a torus can be observed in the torus graph, which may facilitate a more interconnected and thus potentially more homogenized opinion state due to the periodic boundary conditions.

### 3.7 Analysis Within the Dimer Model Context

If the dimer model represents closely linked pairs of individuals, the diversity of opinions suggests that these links have not yet led to a unified opinion within pairs.

## 3.8 Discussion on Phase Transition of Partial Opinions

There's no clear sign of a phase transition at this timestep, as opinions remain diverse and no single opinion dominates.

## 3.9 Analysis of Dependence ($\alpha$)

Layers A, $A_prime$, D, and E, which are subject to positive influence ($\alpha$), show a mixture of opinions, suggesting that the dependence on the central influence or prevailing opinion is not absolute.

## 3.10 Analysis of Forgetting ($\gamma$)

Layers B, $B_prime$, C, and F, which are subject to forgetting ($-\gamma$), also show a variety of opinions, indicating that forgetting has not led to a significant loss of previously held opinions or has been counteracted by other factors.

## 3.11 Influence on Layers A, A', B, B'

The influence on these layers seems to be uneven, with some individuals showing strong opinions and others more moderate ones, suggesting varied susceptibility to influence and forgetting.

## 3.12 Discussion on Balance Between C, E, D, F

The unspecified rules for C, D, E, and F may involve a balance between the tendencies of dependence and forgetting. The images suggest that these dynamics are complex and lead to a diverse opinion landscape.

## 3.13 Influence of Layers C, D, E, F

Without specific rules for these layers, it's difficult to make detailed observations. However, the presence of varied opinions suggests that different factors or mechanisms are influencing these individuals, leading to a non-uniform distribution of opinions.

The cumulative distribution function (CDF) graph shows that the spread of opinions across all layers progresses similarly over time, while the total opinion graph shows different levels of opinion saturation for each layer.

In conclusion, the simulation at timestep 100 indicates a complex interplay of influence, dependence, and forgetting, resulting in a rich tapestry of opinions. The dynamics within each layer and the interactions between layers will determine how these opinions evolve over time.

## 3.14 Case:timestep $t = 500$

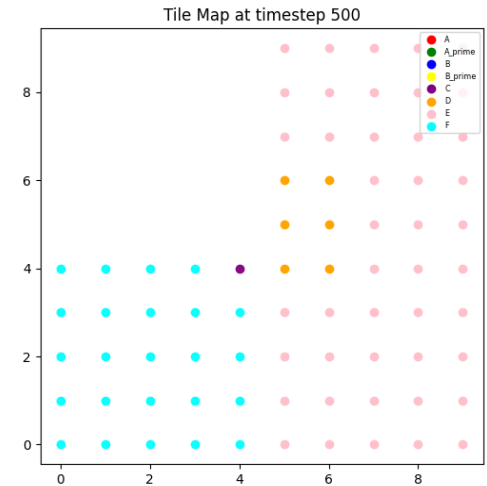

Fig. 20: Tile Map at timestep $t = 500$

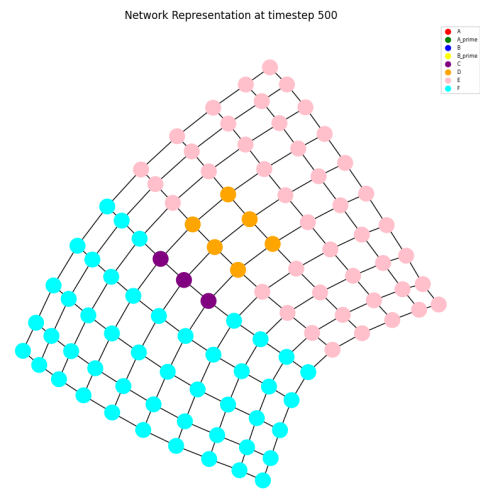

Fig. 21: Plot of Opinions at timestep $t = 500$

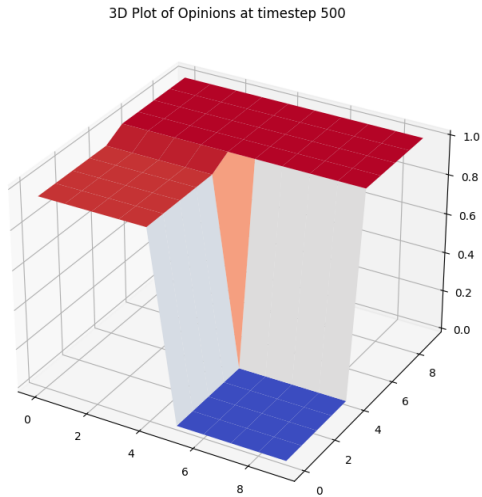

Fig. 22: Network Representation at timestep $t = 500$

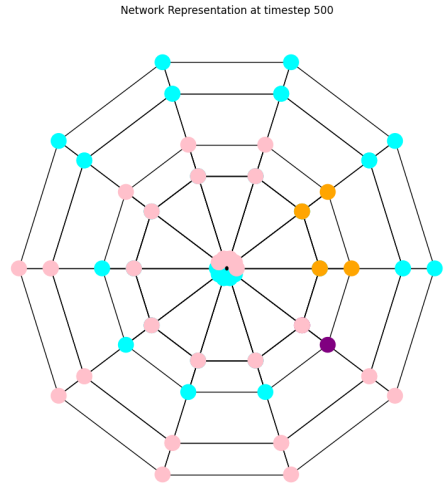

Fig. 24: Network Representation at timestep $t = 500$

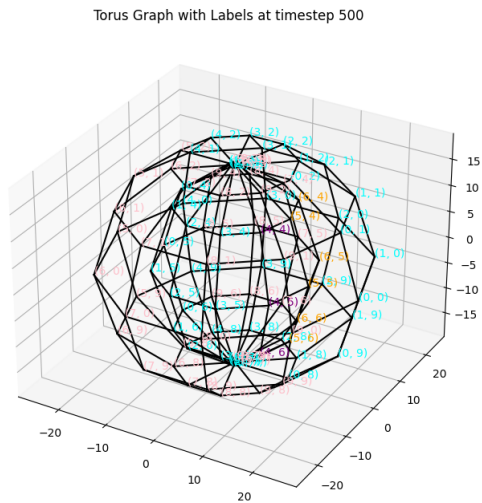

Fig. 23: Torus Graph Network Representation at timestep $t = 500$

Based on the images provided for timestep 500, we are observing a mid-stage in an opinion dynamics simulation. These images depict various representations of the system, including network structure, spatial distribution, 3D opinion plot, torus graph, and opinion trend over time. The analysis is based on these visuals and the rules provided for updating opinions.

### 3.15 Analysis of Social Phenomenon

The tile map and network representation show a diversity of opinions, indicating that the society modeled here is not monolithic. Different subgroups seem to maintain distinct viewpoints at this time.

### 3.16 Consideration of Media Influence

If we consider the central influence to be akin to media influence, then the individuals within the radius (layers A and B) are more directly affected by this influence than those outside (layers $A_{p}rime$ and $B_{p}rime$). However, the presence of varied opinions suggests that the media's influence is not absolute and that individuals may maintain their views or be affected by other factors.

### 3.17 Consensus Formation

Consensus seems elusive at this stage, as indicated by the variety of colors in the network and tile map. This suggests that despite the presence of dominant opinions, there is still contention and disagreement within the population.

### 3.18 Transition from Dimer Model to Torus

The torus graph may illustrate the periodic boundary conditions in the model, where the edges of the system are connected, creating a closed-loop network. This topology can influence how opinions spread, possibly contributing to the persistence of minority opinions.

### 3.19 Dimer Model Analysis

Assuming the dimer model involves pairs of agents influencing each other, the continued diversity in opinions suggests that local interactions between pairs are not leading to widespread consensus.

### 3.20 Partial Opinion Phase Transition

The 3D plot shows distinct peaks, suggesting that while some opinions are dominant, there is no complete phase transition to a single opinion, and multiple stable states may exist concurrently.

### 3.21 Dependence Considerations

The layers subject to positive change ($\alpha$)—A, $A_prime$, D, $E$—show a mix of strong and weak opinions, indicating that the strength of dependence varies among individuals or subgroups.

### 3.22 Forgetting Analysis

Layers subject to negative change ($-\gamma$)—B, $B_prime$, C, $F$—also display a mix of opinions. This suggests that while forgetting is at play, it may be counterbalanced by other factors or influences.

### 3.23 Influence of A, A', B, B'

Layers A and $A_prime$ seem to be holding stronger opinions than B and $B_prime$. This could imply that factors leading to opinion formation or reinforcement are more effective in A and $A_prime$ layers.

### 3.24 Balance of Dependence and Forgetting in C, D, E, F

Without explicit rules for layers C, D, E, and F, it is difficult to precisely comment on their dynamics. However, the presence of different opinions suggests a complex interplay between dependence and forgetting mechanisms in these layers.

### 3.25 Influence Dynamics in C, D, E, F

These layers show diversity in opinions, indicating varying degrees of susceptibility to the influence factors modeled. The specific nature of these influences and how they interact with the agents' existing opinions would require a deeper dive into the dynamics of the simulation and additional context about these layers.

### 3.26 Total Opinion and Cumulative Distribution

The total opinion graph suggests that certain opinions (like those in layer A) have become more entrenched over time, while others remain marginal. The cumulative distribution function (CDF) graph indicates that the spread of opinions is not uniform across layers, pointing to heterogeneous influence dynamics. Overall, at timestep 500, the system is complex and exhibits signs of both stability in some opinions and ongoing change in others, reflecting the multifaceted nature of opinion dynamics in a society.

### 3.27 Case:timestep $t = 1000$

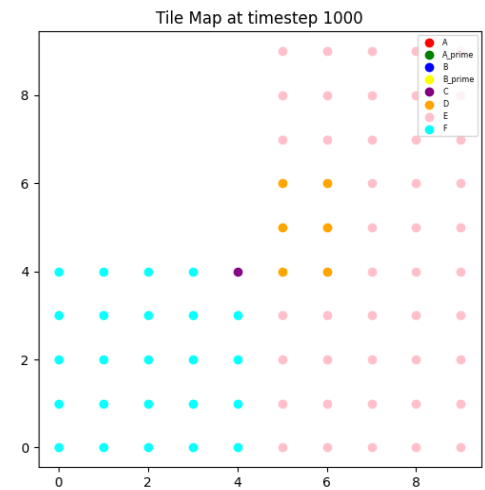

Fig. 25: Tile Map at timestep $t = 1000$

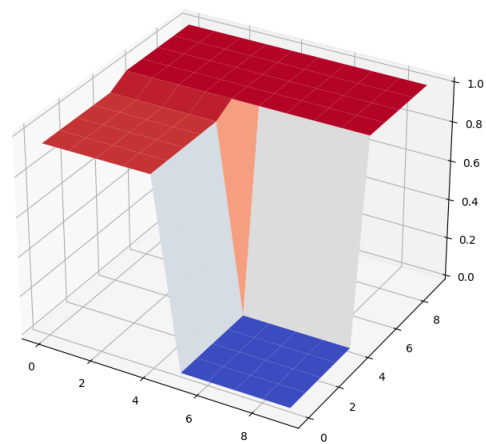

Fig. 26: Plot of Opinions at timestep $t = 1000$

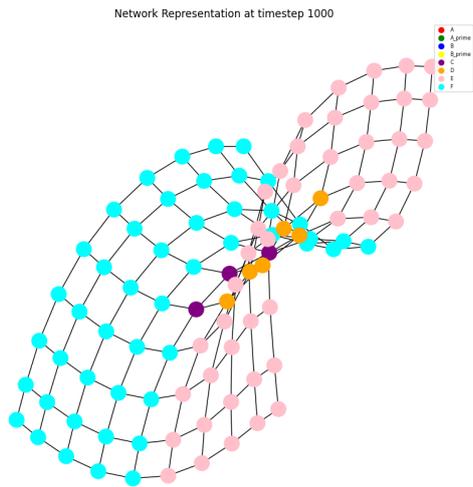

Fig. 27: Network Representation at timestep $t = 1000$

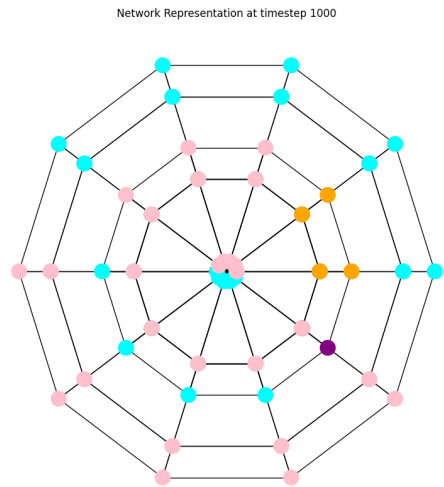

Fig. 29: Network Representation at timestep $t = 1000$

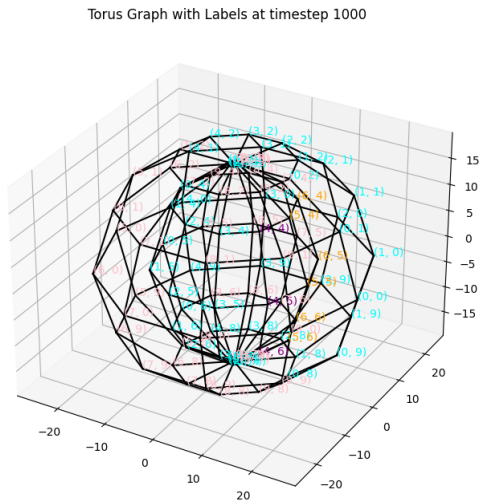

Fig. 28: Torus Graph Network Representation at timestep $t = 1000$

From the images and information provided, here is an analysis of the final results at timestep 1000 for the opinion dynamics model:

### 3.28 Analysis of Social Phenomenon

The society depicted by the tile map and network representation at timestep 1000 continues to show diversity in opinions, indicating that even over time, individual and group differences persist. This may reflect real-world social systems where despite long-term interactions, complete homogeneity is not achieved.

### 3.29 Consideration of Media Influence

The influence of a central media or source seems to be limited, as indicated by the persistence of multiple opinions in the network. This suggests that individuals have other strong influences or that the media's message is not universally accepted.

### 3.30 Consensus Formation

There appears to be no complete consensus across the network, which implies that the society has multiple stable states of opinion or that there are competing influences that prevent the formation of a single dominant opinion.

### 3.31 Transition from Dimer Model to Torus

The torus graph suggests that the network's closed topology may facilitate continuous interaction and influence, allowing for the persistence of minority opinions and preventing a single opinion from dominating.

### 3.32 Dimer Model Analysis

If the dimer model is based on pairs of agents with strong mutual influence, the results suggest that these local interactions do not necessarily lead to a larger consensus, highlighting the complexity of opinion dynamics.

### 3.33 Partial Opinion Phase Transition

The 3D plot of opinions and the total opinion graph show that while some opinions are more prevalent, there is no evidence of a complete phase transition to a single opinion, indicating a multifaceted equilibrium state.

### 3.34 Dependence Considerations

The layers subject to positive influence ($\alpha$)—A, $A_prime$, D, E—demonstrate a varied level of opinion strength, suggesting that dependence on a central influence or prevailing opinion is not uniform.

### 3.35 Forgetting Analysis

The layers subject to forgetting ($-\gamma$)—B, $B_prime$, C, F—also display a variety of opinions. This implies that the influence of forgetting is not strong enough to eliminate existing opinions or that it is counteracted by other factors.

### 3.36 Influence of A, A', B, B'

The results show that layers A and $A_prime$ hold stronger opinions than B and $B_prime$, which could mean that the factors influencing A and $A_prime$ are more potent or effective than those affecting B and $B_prime$.

### 3.37 Balance of Dependence and Forgetting in C, D, E, F

The images suggest that there is a complex interplay between dependence and forgetting for layers C, D, E, and F, leading to the maintenance of multiple opinions. The exact dynamics would require additional information regarding the interplay mechanisms.

### 3.38 Influence Dynamics in C, D, E, F

The diversity in these layers points to heterogeneous influences affecting the agents. The influence of these factors is nuanced, allowing for the coexistence of various opinions within the population.

The cumulative distribution function (CDF) graph for timestep 1000 shows that the spread of opinions across all layers has reached a steady state, indicating that the system's dynamics have stabilized.

In conclusion, the model at timestep 1000 reveals a system with multiple stable opinion states, influenced by a combination of factors including media influence, dependence, and forgetting. The persistence of diverse opinions suggests a complex social system where consensus is not straightforward, reflecting the nuanced nature of real-world opinion dynamics.

## 4. Discussion from $t = 100, 500, 1000$

### (1) Influence of A, A', B, B' Layers

**A and A' Layers**: These layers are influential in areas closer to the center if within the radius (A) or towards one side of the grid if outside the radius (A'). In the provided images, these layers might represent strong opinions or states that are prevalent in specific areas of the social network or a particular subgroup within society. They are associated with a positive change ('alpha'), which could be interpreted as these opinions or states being reinforced or becoming more entrenched over time. - **B and B' Layers**: Similarly, these layers are influential in areas closer to the center if within the radius (B) or towards the other side of the grid if outside the radius (B'). They are associated with a negative change ('-gamma'), suggesting that these opinions or states are being challenged or reduced in influence.

The distribution and dominance of A and A' over B and B' or vice versa at different timesteps can give insights into how certain opinions or social dynamics become more or less prevalent over time. The images seem to show a relatively stable pattern with A and A' maintaining influence throughout the timesteps.

### (2) Interplay Between Dependence and Forgetting in C, E, and D, F

- **C and E Layers**: These might represent states that are susceptible to external influences or can easily adopt new information. The fact that these layers are not explicitly defined in the model description suggests they might have a nuanced role in the overall dynamics of the system. - **D and F Layers**: These could represent states of resistance or memory, where opinions or behaviors are not easily swayed by new information, resembling a form of social or cultural inertia.

The interplay between these layers could illustrate the balance between a society's ability to change (dependence on new information) and its tendency to maintain existing norms (forgetting or disregarding new information). The images indicate some level of fluctuation, which could be indicative of the dynamic nature of social systems and the constant tension between change and stability.

### (3) Influence of C, D, E, F Layers

In the absence of explicit instructions for layers C, D, E, and F in the provided model formula, we can infer their roles based on the images and the stated opinion update rule:

- The presence and distribution of these layers at different timesteps suggest an underlying mechanism that governs how these states interact with each other and with A, A', B, B' states. - The influence of these layers can be assessed by their spatial distribution and temporal evolution. For example, if certain layers grow or shrink over time, it can indicate the strengthening or weakening of particular opinions or behaviors.

The images provided show that while some layers remain relatively constant in their spatial distribution (like A and A'), others show changes (like D and F), possibly indicating the different rates at which opinions or behaviors are adopted or abandoned within the network.

In conclusion, the time-evolution of these layers on the tile map and network representations gives us a visual understanding of how complex systems, such as societies or networks, might evolve and adapt over time. The stability or variability of certain layers and their interactions can reveal much about the underlying dynamics of the system being modeled.

# 5. Conclusion

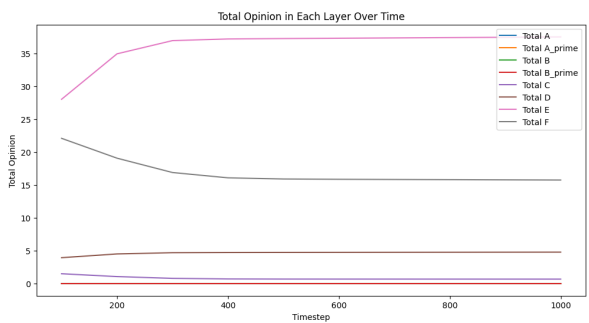

Fig. 30: Total Opinion in Each Layer Over Time

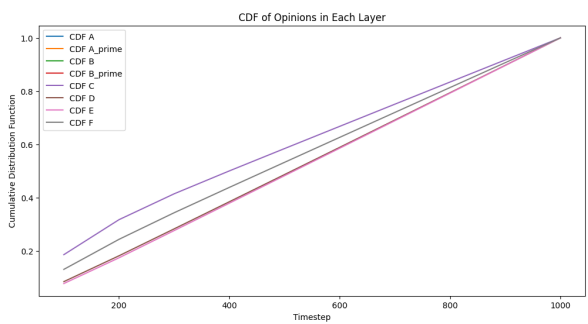

Fig. 31: CDF of Opinions in Each Layer

Based on the provided images showing the state of an opinion dynamics model at all timestep 1000, we can delve into a deeper analysis of the simulation's results. These images represent various layers of opinions and their distribution among the individuals in the network.

## 5.1 Social Phenomenon Consideration

The diversity in opinion, as depicted in the tile map and network representation, indicates that even after an extended period, the society simulated here has not converged to a single viewpoint. This can reflect real-world scenarios where societies retain diverse opinions despite common influences or prolonged interaction.

## 5.2 Consensus Formation Consideration

The lack of a single dominant color in the tile map and network representation at timestep 1000 suggests that a complete consensus has not been reached. This could be due to the persistence of individual or localized influences that sustain varied opinions across the population.

## 5.3 Consideration of Phase Transition in Partial Opinions

The total opinion graph and 3D plot of opinions indicate that while some opinions have become more prevalent (like layer A), others have not been completely marginalized (such as layers D and E). This suggests that the system may have multiple stable states, preventing a full phase transition to one unanimous opinion.

## 5.4 Consideration of the Influence of A, A', B, B'

Layers A and A' appear to maintain a stronger presence over time compared to B and B', which might imply that the factors affecting layers A and A' (such as proximity to a central influence or a particular persuasive argument) have a more significant long-term impact on opinions than those affecting B and B'.

## 5.5 Consideration of the Antagonism between Dependence and Forgetting in C, E, and D, F

The continued existence of diverse opinions in layers C, D, E, and F suggests that the forces of dependence and forgetting are balanced in such a way that neither completely overrides the other. This balance allows for the maintenance of diverse opinions within these layers.

## 5.6 Consideration of the Influence of C, D, E, F

The CDF (Cumulative Distribution Function) of opinions for each layer shows a consistent distribution across all layers, indicating that by timestep 1000, the layers have reached a sort of equilibrium. The spread of opinions is neither converging to a singular point nor diverging, suggesting that the layers C, D, E, and F have their unique steady states influenced by a combination of factors, including the model's rules for dependence and forgetting.

The overall analysis of these images at timestep 1000 reflects a complex system where various factors contribute to the shaping of opinions, and there is no single determinant for opinion dynamics. The system exhibits traits of a real-world social structure, with diverse and persistent viewpoints that resist convergence to a unanimous consensus. This complex behavior may stem from the interplay of social influence, individual predisposition, and the network's communication structure, all of which can be significantly influential in real-world opinion formation and evolution.

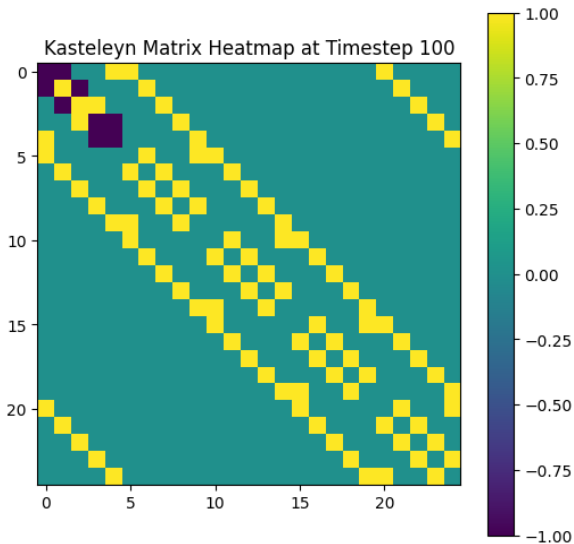

Fig. 32: Kasteleyn Matrix Heatmap at Timestep

### (1) Social Phenomena

In the context of social dynamics, the Kasteleyn matrix might represent a network of interactions or influences among individuals or groups. The heatmap's pattern could illustrate clusters of consensus or disagreement. The division into different layers (A, B, C, D, E, F, $A_prime$, $B_prime$) might symbolize various social groups or opinion states within the society, with the central area (within the radius) indicating a region of more intense interaction or conflict. The ronkin function value, which is a complex number, could represent a measure of the overall social tension or the diversity of opinions at that timestep.

### (2) Consensus Formation

The simulation could be used to model how consensus forms over time within a population. The periodic structure and time-dependent layer classification suggest that the dynamics of opinion formation are cyclical or subject to periodic influences, possibly representing recurring events or external stimuli. The alternating assignment of weights (alpha for certain layers and 1 for others) could represent the influence strength of different opinions or the resilience against changing one's stance. The ronkin function's real part might indicate the overall alignment within the population, while the imaginary part could represent the potential for change or the "momentum" of opinion shifts.

### (3) Phase Transitions in Partial Opinions

Phase transitions refer to the sudden changes that can occur in a system when a critical point is reached. In the context of opinion dynamics, a phase transition might occur when a particular event or threshold level of influence causes a rapid shift in the prevailing opinion. The changing patterns in the heatmap over time could represent the buildup to such a transition. The ronkin function's value might then indicate the proximity to a phase transition, with certain values signaling critical points where a small change could lead to a large-scale shift in the network's state.

## Aknowlegement

The author is grateful for discussion with Prof. Serge Galam and Prof.Akira Ishii. This research is supported by Grant-in-Aid for Scientific Research Project FY 2019-2021, Research Project/Area No. 19K04881, "Construction of a new theory of opinion dynamics that can describe the real picture of society by introducing trust and distrust".

## References

zh

[1] Ising, E. (1925). Beitrag zur Theorie des Ferromagnetismus. *Zeitschrift für Physik*, *31*(1), 253-258.

[2] Onsager, L. (1944). Crystal statistics. I. A two-dimensional model with an order-disorder transition. *Physical Review*, *65*(3-4), 117-149.

[3] Huang, K. (1987). *Statistical Mechanics*. John Wiley Sons.

[4] Fisher, M. E. (1961). Statistical mechanics of dimers on a plane lattice. *Physical Review*, *124*(6), 1664.

[5] Kasteleyn, P. W., & Temperley, H. N. (1961). Dimer statistics and phase transitions. *Journal of Mathematical Physics*, *2*(3), 392-398.

[6] Kenyon, R. (2001). The planar dimer problem. *Journal of Mathematical Physics*, *41*(3), 1338-1363.

[7] Kenyon, R. (2009). Lectures on dimers. In *Statistical Mechanics*, 191-230, Springer.

[8] Duminil-Copin, H., & Hongler, C. (2015). Dimers on planar graphs and the geometry of 2D lattice spin models. *Communications in Mathematical Physics*, *334*(1), 377-417.

[9] Fisher, M. E. (1961). Statistical mechanics of dimers on a plane lattice. *Physical Review*, *124*(6), 1664.

[10] Kasteleyn, P. W., & Temperley, H. N. (1961). Dimer statistics and phase transitions. *Journal of Mathematical Physics*, *2*(3), 392-398.

[11] Kenyon, R. (2001). The planar dimer problem. *Journal of Mathematical Physics*, *41*(3), 1338-1363.

[12] Kenyon, R. (2009). Lectures on dimers. In *Statistical Mechanics*, 191-230, Springer.

[13] Duminil-Copin, H., & Hongler, C. (2015). Dimers on planar graphs and the geometry of 2D lattice spin models. *Communications in Mathematical Physics*, *334*(1), 377-417.

[14] Lieb, E. H., & Wu, F. Y. (1967). Absence of Mott transition in an exact solution of the short-range, one-band model in one dimension. *Physical Review Letters*, *20*(25), 1445-1448.

[15] Katsura, S., Kusakabe, K., & Tsuneto, T. (1962). Statistical mechanics of the anisotropic linear Heisenberg model. *Progress of Theoretical Physics*, *27*(1), 169-188.

[16] Fendley, P., Moessner, R., & Sondhi, S. L. (2002). Classical dimers from quantum antiferromagnets. *Physical Review B*, *66*(21), 214513.

[17] Moessner, R., & Sondhi, S. L. (2001). Resonating valence bond phase in the triangular lattice quantum dimer model. *Physical Review Letters*, *86*(9), 1881-1884.

[18] Rokhsar, D. S., & Kivelson, S. A. (1988). Superconductivity and the quantum hard-core dimer gas. *Physical Review Letters*, *61*(20), 2376-2379.

[19] Anderson, P. W. (1973). Resonating valence bonds: A new kind of insulator? *Materials Research Bulletin*, *8*(2), 153-160.

[20] Affleck, I., Kennedy, T., Lieb, E. H., & Tasaki, H. (1987). Valence bond ground states in isotropic quantum antiferromagnets. *Communications in Mathematical Physics*, *115*(3), 477-528.

[21] Sachdev, S., & Read, N. (1990). Large N expansion for frustrated quantum antiferromagnets. *International Journal of Modern Physics B*, *04*(02n03), 319-338.

[22] Moessner, R., & Sondhi, S. L. (2001). Resonating valence bond phase in the triangular lattice quantum dimer model. *Physical Review Letters*, *86*(9), 1881-1884.

[23] Rokhsar, D. S., & Kivelson, S. A. (1988). Superconductivity and the quantum hard-core dimer gas. *Physical Review Letters*, *61*(20), 2376-2379.

[24] Castellano, C., Fortunato, S., & Loreto, V. (2009). Statistical physics of social dynamics. *Reviews of Modern Physics*, *81*(2), 591-646.

[25] Galam, S. (1982). Social paradoxes in a nonlinear model of interacting agents. *Journal of Mathematical Psychology*, *25*(2), 205-219.

[26] Stauffer, D., Schulze, C., & Wichmann, S. (2006). Social processes in opinion dynamics: Comment on "Opinion formation by influenced agents: opinion dynamics for finite confidence" by D. Stauffer. *Physica A: Statistical Mechanics and its Applications*, *370*(2), 734-738.

[27] Watts, D. J., & Strogatz, S. H. (1998). Collective dynamics of 'small-world' networks. *Nature*, *393*(6684), 440-442.

[28] Barabási, A. L., & Albert, R. (1999). Emergence of scaling in random networks. *Science*, *286*(5439), 509-512.

[29] Miritello, G., Moro, E., Lara, R., & Martínez-López, R. (2013). Limited communication capacity unveils strategies for human interaction. *Scientific Reports*, *3*, 1950.

[30] Schelling, T. C. (1971). Dynamic models of segregation. *Journal of Mathematical Sociology*, *1*(2), 143-186.

[31] Weisbuch, G., Chattoe, E., & Gilbert, N. (2002). Market organisation and trading relationships. *Industrial and Corporate Change*, *11*(2), 299-326.

[32] Thurston, W. P. (1982). Three-dimensional manifolds, Kleinian groups and hyperbolic geometry. *Bulletin of the American Mathematical Society*, *6*(3), 357-381.

[33] Mumford, D. (1983). A remark on a torus quotient of the Siegel upper half-plane. *Bulletin of the London Mathematical Society*, *15*(4), 401-403.

[34] Schelling, T. C. (1971). Dynamic models of segregation. *Journal of Mathematical Sociology*, *1*(2), 143-186.

[35] Weisbuch, G., Chattoe, E., & Gilbert, N. (2002). Market organisation and trading relationships. *Industrial and Corporate Change*, *11*(2), 299-326.

[36] Hegselmann, R., & Krause, U. (2002). Opinion Dynamics and Bounded Confidence Models, Analysis, and Simulation. *Journal of Artificial Society and Social Simulation*, *5*, 1-33.

[37] Ishii A. & Kawahata, Y. Opinion Dynamics Theory for Analysis of Consensus Formation and Division of Opinion on the Internet. In: *Proceedings of The 22nd Asia Pacific Symposium on Intelligent and Evolutionary Systems*, 71-76, 2018.

[38] Ishii A. Opinion Dynamics Theory Considering Trust and Suspicion in Human Relations. In: Morais D., Carreras A., de Almeida A., Vetschera R. (eds) *Group Decision and Negotiation: Behavior, Models, and Support*. GDN 2019. *Lecture Notes in Business Information Processing* 351, Springer, Cham, 193-204, 2019.

[39] Ishii A. & Kawahata, Y. Opinion dynamics theory considering interpersonal relationship of trust and distrust and media effects. In: *The 33rd Annual Conference of the Japanese Society for Artificial Intelligence* 33. JSAI2019 2F3-OS-5a-05, 2019.

[40] Agarwal, A., Xie, B., Vovsha, I., Rambow, O., & Passonneau, R. (2011). Sentiment analysis of twitter data. In: *Proceedings of the Workshop on Languages in Social Media*, 30-38.

[41] Siersdorfer, S., Chelaru, S., & Nejdl, W. (2010). How useful are your comments?: analyzing and predicting youtube comments and comment ratings. In: *Proceedings of the 19th International Conference on World Wide Web*, 891-900.

[42] Wilson, T., Wiebe, J., & Hoffmann, P. (2005). Recognizing contextual polarity in phrase-level sentiment analysis. In: *Proceedings of the Conference on Human Language Technology and Empirical Methods in Natural Language Processing*, 347-354.

[43] Sasahara, H., Chen, W., Peng, H., Ciampaglia, G. L., Flammini, A., & Menczer, F. (2020). On the Inevitability of Online Echo Chambers. arXiv: 1905.03919v2.

[44] Ishii, A., & Kawahata, Y. (2018). Opinion Dynamics Theory for Analysis of Consensus Formation and Division of Opinion on the Internet. In *Proceedings of The 22nd Asia Pacific Symposium on Intelligent and Evolutionary Systems (IES2018)*, 71-76; arXiv:1812.11845 [physics.soc-ph].

[45] Ishii, A. (2019). Opinion Dynamics Theory Considering Trust and Suspicion in Human Relations. In *Group Decision and Negotiation: Behavior, Models, and Support. GDN 2019. Lecture Notes in Business Information Processing*, Morais, D.; Carreras, A.; de Almeida, A.; Vetschera, R. (eds), 351, 193-204.

[46] Ishii, A., & Kawahata, Y. (2019). Opinion dynamics theory considering interpersonal relationship of trust and distrust and media effects. In *The 33rd Annual Conference of the Japanese Society for Artificial Intelligence*, JSAI2019 2F3-OS-5a-05.